\documentstyle[preprint,aps,epsf,floats]{revtex}

\begin{document}
\tighten
\newcommand{\pislash}{ {\pi\hskip-0.6em /} }
\newcommand{\pislashsmall}{ {\pi\hskip-0.375em /} }
\newcommand{\nopi}{{\rm EFT}(\pislash) }
\baselineskip=18.0pt
\parskip=12.0pt
\parindent=2em

\title{EFFECTIVE FIELD THEORY IN NUCLEAR PHYSICS}

\author{{\bf Martin J. Savage}$^{a,b}$~\footnote{NT@UW-00-036}
  and {\bf Barry R. Holstein}$^c$}
\address{$^a$ Department of Physics, University of Washington, \\
Seattle, WA 98195. }
\address{$^b$ Jefferson Laboratory, 12000 Jefferson Avenue,\\
Newport News, VA 23606.}
\address{$^c$ Department of Physics,
University of Massachusetts,\\
Amherst, MA 01003.}
\maketitle

\begin{abstract}
The Electromagnetic and Hadronic physics sub-community of nuclear physics
held a town hall meeting at Jefferson Lab during November 30 to December 4
of 2000.
This is is our combined contribution to the white paper that 
will result from this meeting.
\end{abstract}

\vskip -0.5in

\section{The fundamental scientific questions addressed}

A fundamental challenge facing the nuclear physics community is the
development of a theory that describes the properties and dynamics of
the strongly interacting mesons and baryons.  This theory must be
consistent with QCD for processes below the chiral symmetry breaking
scale, a scale comparable with the nucleon mass.  The importance of
meeting this challenge cannot be overstated as it is vital for a
theoretical description of nuclei and multi-nucleon processes,
including electroweak probes.  Further, it is necessary in order to
make reliable predictions in conditions that are not accessible to
controlled measurement, such as those that exist in supernovae, or
during hadronization following the formation of a quark-gluon plasma.

Effective field theory (EFT) organizes quantum field theories
according to hierarchies of physical scales~\cite{Ma96}.
The Standard Model of electroweak interactions is a
beautiful example of an EFT, which describes observables at
energies below the scale of electroweak symmetry breaking. Its
reliability and rigor are undisputed.  An EFT is the most general
description that exists consistent with all underlying symmetries and
physical principles.
The uncertainty  associated with a calculation of any observable in an
EFT can be estimated and controlled. It is sometimes the case that in
a particular limit of the parameter space of the underlying theory,
additional symmetries become manifest, e.g. chiral and heavy-quark
symmetries in QCD.  The EFT then allows one to calculate
perturbatively about the symmetry limit.  To describe hadronic
interactions, a dual expansion in the up and down quark masses, $m_u$
and $m_d$, and in the momentum of external probes is required. This
approach was pioneered by Weinberg~\cite{We67} and has been
successfully applied in the meson sector including both two- and
three-flavors~\cite{Ga82}, and in the single nucleon sector (for a
recent review see Ref.~\cite{Me00}).  This body of work, collectively
known as chiral perturbation theory ($\chi$PT),
provides a cornerstone in our
understanding of QCD and is the only rigorous way in which to encode
the entire body of QCD predictions at low energy.  Nevertheless,
challenges remain in its development, as will be discussed below.

The last decade has also seen important
progress in the extension of the
hadronic EFT to systems involving more than one nucleon,
that is to say, toward the construction of
nuclear chiral perturbation theory (N$\chi$PT).
The
intrinsically nonperturbative nature of nuclei renders this endeavor
highly nontrivial. Influential work by Weinberg~\cite{We90} in the
early 1990's and by many others (for a review and extensive
referencing for what follows see Ref.~\cite{Be00}) in the late
1990's has led to remarkable progress in describing multi-nucleon
systems at low energies. A few samples will be discussed below. There
are, however, many unresolved issues about how best to organize and
optimize the EFT expansion, in particular as regards the treatment of
the pion. Resolution of these issues is essential in order to develop
a systematic expansion about the chiral limit and to push the range of
validity of the EFT beyond the Fermi momentum of nuclear matter.

\section{ Major achievements since the last Long Range Plan}

\subsection{Chiral Perturbation Theory}

Chiral perturbation theory was discovered by
Weinberg\cite{We67} in the late 1960's and 1970's.
Subsequently, it was developed in the early 1980's by Gasser and
Leutwyler\cite{Ga82} who wrote down the most general
counterterm Lagrangian for mesons
at one-loop order (${\cal O}(p^4)$)
including
ten {\it apriori} unknown parameters, commonly known as the $L_i$'s.
By comparing the one-loop order predictions with experiment,
empirical values for the $L_i$ can be obtained.
Given the simplicity of the theory at one-loop order,
it is quite predictive, despite the appearance of the $L_i$'s.
Examples are shown in Table 1, where predictions are
compared with experimental determinations for quantities
that receive contributions from just two of the
constants---$L_9,L_{10}$.
The table reveals at least one intriguing
problem---the solid prediction of $\chi$PT
for the electric polarizability of the $\pi^+$
may be violated.
However, of the three experimental results only one
can be considered to be in disagreement.
Clearing up this discrepancy should be a focus of
future experimental work in this area.
\begin{table}[!ht]
\begin{center}
\begin{tabular}{cccc}\hline\hline
Reaction&Quantity&Theory&Experiment\\
\hline
$\pi^+\rightarrow e^+\nu_e\gamma$ & $h_V(m_\pi^{-1})$ & 0.027
& $0.029\pm 0.017$~\cite{Pd96}\\
$\pi^+\rightarrow e^+\nu_ee^+e^-$ & $r_V/h_V$ & 2.6 & $2.3\pm
0.6$~\cite{Pd96}\\
$\gamma\pi^+\rightarrow\gamma\pi^+$ & $(\alpha_E+\beta_M)\,(10^{-4}\,{\rm
fm}^3)$& 0
&$1.4\pm 3.1$~\cite{An85}\\
      &$\alpha_E\,(10^{-4}\,{\rm fm}^3)$&2.8 & $6.8\pm 1.4$~\cite{An83}\\
 & & & $12\pm 20$~\cite{Ai86}\\
 & & & $2.1\pm 1.1$~\cite{Ba92}\\
\hline
\end{tabular}
\caption{Predictions of $\chi$PT
and data for radiative pion processes.}
\end{center}
\end{table}

\vskip -0.2in

$\chi$PT has been extended to include baryons
where it is found that in order to have a consistent
power-counting it is necessary to perform a simultaneous expansion in
energy-momentum and in
inverse powers of the nucleon mass~\cite{JM91}.
This procedure is called heavy
baryon chiral perturbation theory (HB$\chi$PT)~\cite{JM91}
and has been used  to
address most of the problems in low energy baryonic
interactions~\cite{Me00}.
As opposed to the case of the mesons discussed above, however, there
are issues which are still not completely understood.
The convergence of the perturbative expansion
is slower than in the meson sector and
in some cases chiral SU(3) loops
produce large effects which must be partially
canceled by the corresponding counterterms.
Both these problems are under study.
However, there are a large number of successful predictions with
which to challenge experiment, especially in the realm of Compton
scattering, both real and virtual.
In many cases, calculations and
empirical numbers are available for the polarizabilities and
generalized polarizabilities\cite{He00}, which characterize the
response of a system to an applied electromagnetic field.
Also,
within the last few years precision experimental
results for near threshold pion photoproduction and electroproduction
have become available.
Theoretical calculations at ${\cal O}(p^4)$
generally compare well with data but convergence may be
a problem for the S-wave multipoles.

In order to make a connection between the numerical results
of present day lattice-QCD efforts
(partially-quenched and with unphysical quark masses)
and nature, the particle physics community
has developed an effective field theory tool~\cite{Sh92}.
This tool is known as partially-quenched-$\chi$PT (PQ$\chi$PT),
and is being used successfully in the particle physics
lattice-QCD community.

In general, a reliable calculational framework in
both the meson and single nucleon sectors exists with which to confront
precise experimental data.
However, important theoretical challenges remain:
\begin{itemize}
\item [i)] Understanding how to deal with convergence issues,
especially in the nucleon sector.
\item [ii)] Understanding how to take this reliable calculational
formalism to higher energies than are presently possible.
\item [iii)] Extending existing calculations to sectors
wherein present calculations are lacking, e.g. processes involving the
$\eta$ and $\eta^\prime$.
\item [iv)] Understanding how to calculate the empirical constants
from first principles, rather than dealing with them
phenomenologically.
\end{itemize}
Work is underway on these issues at the present time:
\begin{itemize}
\item [i)] Extending specific calculations to two-loop order\cite{Ga99}.
\item [ii)] Attempts to
marry chiral and dispersive methods to achieve
unitary amplitudes which are valid to much higher energies than
previously thought possible~\cite{Do93}.
\item [iii)] Calculations involving both the $\eta$ and $\eta'$\cite{Ka96}.
\item [iv)] Analytical techniques combined with lattice-QCD efforts
in the particle theory community
have begun to confront the empirical values of the $L_i$'s with underlying
theory.
\item [v)] Work which marries chiral and $1/N_c$ methods\cite{Fl00}.
\end{itemize}

\subsection{Multi-Nucleon EFT}

\begin{figure}[!ht]
\vskip 0.15in
\centerline{{\epsfxsize=6in \epsfbox{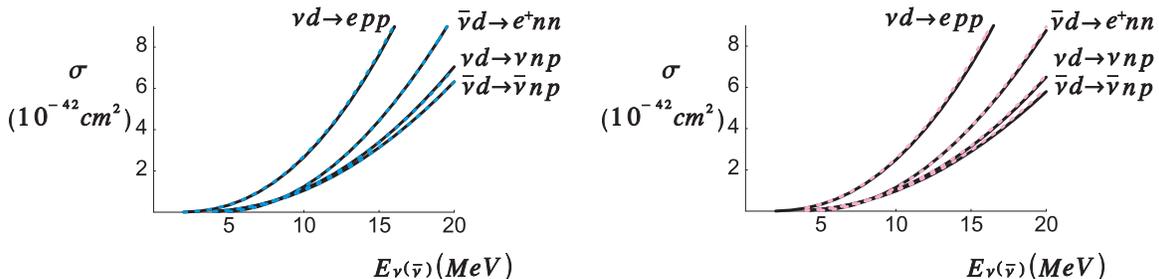}} }
\caption{
Inelastic $\nu (\overline{\nu}) d$ cross sections as a function
of incident $\nu (\overline{\nu})$ energy.
In the left panel the solid curve is the result of
the potential model calculation of
Ref.~\protect\cite{Na00}
while the dashed curve is the $\nopi$ result with
$L_{1,A}=5.6~{\rm fm}^3$~\protect\cite{Bu00}.
In the right panel the solid curve is the result
the potential model calculation of
of Ref.~\protect\cite{Yi89}
while the dashed curve is the $\nopi$ result with
$L_{1,A}=0.94~{\rm fm}^3$~\protect\cite{Bu00}.
}
\label{fig:neutrino}
\end{figure}

For very low-energy processes, involving energy and momentum much less
than the pion mass, an EFT in the two-nucleon sector has been
developed, $\nopi$, which allows
for a perturbative calculation of processes involving two
nucleons. Consider $np\rightarrow d\gamma$ and $\nu
d$-break-up. First, the cross section, $\sigma_{np}$, for the
radiative capture $np\rightarrow d\gamma$ plays a central role in
predicting the abundance of elements from Big-Bang nucleosynthesis
(BBN). For many years an error of $5\%$ was assigned to $\sigma_{np}$.
Recently, $\sigma_{np}$ was computed in $\nopi$, and is now described
at the $\sim 1\%$ level by a compact analytic
expression. Second, cross sections for $\nu d$-break-up
are required input in order to determine the flux of neutrinos from
the sun with the Sudbury Neutrino Observatory (SNO).
The differences among existing potential model calculations
are at the $\sim 5\%$-level, arising primarily from differing
treatment of meson-exchange-currents (MEC's).
Using $\nopi$,  it has been
shown that in order to perform a $\sim 1\%$
calculation of this cross section, one {\it a priori} unknown
coefficient, $L_{1,A}$ needs to be determined.  Comparisons between
the analytic EFT calculation and the numerical potential model
calculations are shown in Figure~\ref{fig:neutrino}.
As it is likely to be many years before lattice-QCD can produce a value
for $L_{1,A}$, an experiment to measure the cross section of $\nu_e
d\rightarrow ppe^-$ at the $\sim 1\%$-level is supported.  Such a
measurement will allow for a $\sim 1\%$-level prediction of the other
break-up channels.  In addition to these two processes,
$\nopi$ has been used to make precise predictions for other
two-nucleon observables, such as $\gamma d\rightarrow \gamma d$
Compton scattering that may, with precise experimental measurements at
low-energies, yield reliable determinations of the polarizabilities of
the neutron. Significant progress has also been made in the
computation of three-body scattering cross sections and in our
understanding of how multi-nucleon operators, such as the three-body
force, contribute to low-energy processes.
Spectacular results have been obtained in the spin-${3\over 2}$
channel, where the $nd$ scattering length has been computed at
next-to-next-to-leading order in perturbation theory to be
$a_{3/2}^{\rm EFT} = 6.33 \pm 0.05~{\rm fm}$, which is to
be compared with the experimental determination of $a_{3/2}^{\rm expt}
= 6.35 \pm 0.02~{\rm fm}$ (subsequent second generation potential
model calculations also agree with $a_{3/2}^{\rm expt}$).
The calculation of the energy dependence is similarly impressive.  The
Phillips line, relating the triton binding energy and the three-body
scattering length is recovered, and has been shown to result from the
freedom associated with choice of three-body force.  In
addition, $Nd$ scattering has been studied in the p-wave and
higher partial waves, producing very nice predictions that are yet to be
confirmed experimentally.  The techniques developed for the
three-nucleon systems have been successfully applied to some
observables in the area of Bose-Einstein condensation (BEC).  On a more
technical note, the appearance of a one-parameter limit-cycle in the
renormalization group evolution of the three-body force is of current
interest.

\begin{figure}[!ht]
\vskip 0.15in
\centerline{{\epsfxsize=3in \epsfbox{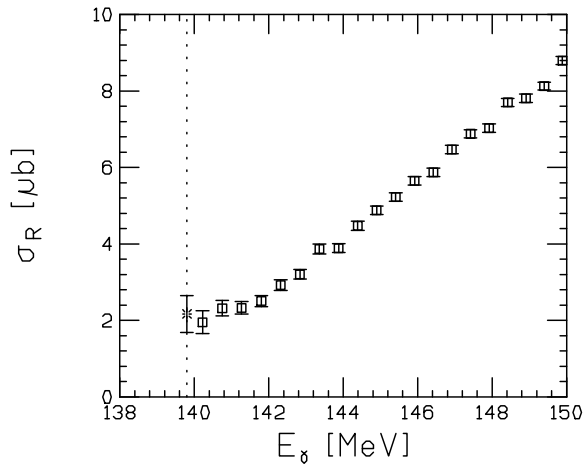}} }
\vskip 0.15in
\caption{The total cross section for $\gamma d \to \pi^0 d$.
Data from
SAL are depicted by the boxes, while
the chiral perturbation theory
threshold prediction and its associated uncertainty
(computed prior to the experiment) is the star and error bar on
the vertical dotted line (threshold photon energy).}
\label{fig:jack}
\end{figure}

The EFT description including pions has also advanced significantly.
Weinberg's original proposal to compute $NN$ potentials using the
organizational principles of $\chi$PT has been
widely developed.  Nucleon-nucleon scattering phase
shifts, $\gamma d$ Compton scattering,
$\gamma d\rightarrow\pi^0 d$ (see Figure~\ref{fig:jack}
for a comparison of the threshold prediction with data), and other
inelastic processes have been computed with great success using
Weinberg power counting. Three-body calculations at third order in
Weinberg power counting are underway, and preliminary results appear
to provide insight into the well-known $A_y$ puzzle.
Weinberg's method is intrinsically numerical and is similar in spirit
to traditional nuclear physics potential theory.  Unfortunately, the
renormalization group scaling of operators in this EFT is complicated.
Moreover, there appear to be inconsistencies in the handling of
divergences proportional to the quark masses, which can
potentially lead to uncontrolled errors.  In contrast,
KSW power counting, where pions contribute at subleading
orders in the expansion, and which allows for analytic results, is
found not to converge at higher orders in the spin-triplet
channels. In exploring these two different types of power
counting a large amount of expertise has been acquired, and efforts
are ongoing to formulate a consistent and converging power counting,
that is sure to involve ingredients from both Weinberg and KSW
power-counting schemes.

Work is underway to incorporate the rigor introduced by the EFT
framework into the nuclear many-body problem.
While this effort is still in its infancy, very
encouraging results have been obtained which suggest that these new
techniques will lead to a dramatic simplification in this arena.

\section{Short Term ($<$3 yrs) and Long Term ($<$10 yrs) Outlook}

There are several lines of investigation that need to be pursued in
order to make substantial progress in low-energy nuclear physics.

The key issue here is how to deal rigorously with the consequences of
QCD in the low energy (nonperturbative) realm.  In the case of mesonic
and single nucleon processes, methods based on the
chiral symmetry of QCD offer perhaps the best way to addresss this
challenge in the near term.  Chiral perturbation theory is a reliable
low energy procedure that is very successful in this regard.
Nevertheless a number of important challenges remain:
\begin{enumerate}

\item[i)] In the realm of pseudoscalar meson interactions, we already have
developed a very successful calculational scheme.  The challenge here
is to experimentally resolve some of the  remaining thorny
problems, such as the pion polarizability, and to extend the predictive
power to higher energy.  Present
theoretical efforts which combine chiral effective theory with other
techniques such as dispersion relations have been able to
substantially increase the energy range over which solid predictions
can be made.

\item[ii)] In the baryonic realm, it is essential not only to extend the
predictive power to higher energy but also to solve the convergence
issues which plague some observables.  The same dispersive methods
which are successful in the mesonic realm may be of utility here, as
may well be new summations of the chiral series which respect the
singularity structure of the basic amplitudes.  Both approaches are
being pursued at the present time.

\item[iii)] Dyson-Schwinger as well as other techniques, such as
quenched lattice-QCD, have been used in
order to compare theoretical and empirical values of the low energy
constants.  Such work will no doubt continue, and represents the
ultimate challenge to any nonperturbative theoretical program.

\end{enumerate}

In the case of multi-nucleon EFT:

\begin{enumerate}
\item[i)] A  consistent and convergent power counting must be established
for systems of nucleons and pions below the chiral symmetry breaking
scale.  This will allow for the calculation of processes that are both
experimentally accessible and inaccessible; high precision
calculations will be performed to make critical comparisons with low-
and moderate-energy nuclear experiments. Such a comparison is vital to
test the convergence of the theory for a variety of processes, but
will also allow for the determination of nucleon properties.
A good
example of this would be the extraction of the nucleon anapole
moment from low-energy electron scattering. Also, precise predictions
of a multi-nucleon EFT will provide a bench mark for lattice-QCD
calculations in much the same way as chiral perturbation theory
presently does in the meson sector.

\item[ii)] Including electroweak gauge fields into systems with three or more
nucleons is a short term priority.  There are large theoretical
uncertainties in potential model calculations of some electromagnetic
processes of great importance in astrophysical environments. It is
essential that these uncertainties be significantly reduced.  Recall
that BBN calculations and interpretation of SNO data are
significantly impacted by uncertainties in
theoretical nuclear physics predictions.  Given the
success in the two-nucleon sector, extension to the three-nucleon
sector is important.

\item[iii)] The early efforts to implement an EFT description of 
multi-nucleon
systems must continue, and be extended to nuclei of moderate atomic
number. The calculations of the Argonne group will provide
a bench mark.  The systematic inclusion of chiral symmetry will allow
for inelastic processes involving pions to be computed in the same
framework as the computation of nuclear energy levels.

\item[iv)] As numerical studies of lattice-QCD will be unable to
directly compute observables in multi-nucleon systems 
in the foreseeable future,
in order for such efforts to have implications that are not purely academic,
a partially-quenched multi-nucleon EFT is required.

\item[v)] High precision calculations continue to be carried out with $\nopi$.
These calculations of low-energy processes can be continued to even
higher orders, allowing for calculations with uncertainties at the
fraction of $1\%$-level.  In conjunction with low-energy experiments
of comparable precision, fundamental properties of the nucleon can be
determined, in addition to allowing for precise predictions for
processes of astrophysical interest.

\item[vi)] Finally, ultimately it must be shown that many-body nuclear
physics methods emerge as a leading order effect in EFT. This will
open the way for systematic improvement of these methods.

\end{enumerate}

\section{Comparison of U.S. and Global Effort}

The development of EFT for particle and nuclear physics is a global effort.
Important groups which focus on chiral perturbative studies---both in
particle and nuclear physics---exist in
England, Germany, Switzerland, Austria, Italy, France, Sweden, Canada, and in
the United States.  Groups which focus on multi-nucleon systems exist
in France, Korea, Germany, Switzerland, England and the United States.
In the latter case, the simultaneous need for knowledge of nuclear
phenomenology and of quantum field theory techniques significantly
restricts the number of physicists who have contributed in any
meaningful way to the development of this area. It is only during the
last few years or so that there has been substantial effort
and progress in this field. Nevertheless, this progress has been impressive.

A good indication of the recent effort in chiral perturbation theory
can be found in the proceedings of the recent Chiral Dynamics meetings
which have taken place at MIT (1994)\cite{Be95}, at Mainz
(1997)\cite{Be98},
and at JLab (2000)\cite{Be00B}.  In each case about 100 physicists,
both theorists and
experimentalists got together to discuss developments in the field.
Discussions were lively and a program setting out future work was developed.
There was a good mix of ($\sim$ 70) senior and ($\sim$ 30) younger
colleagues.  However, the opportunities for tenure track positions has
been quite limited.

An indication of the effort in the area of two-nucleon EFT can be found in the
attendance at the second conference in this area during February of 1999
(proceedings from the first and second
conferences on {\it Nuclear Physics with Effective Field Theory}
can be found in Ref.~\cite{EFTproc}).
Those in attendance at that meeting included twenty
(20) active physicists in tenured or tenure track positions, and
twenty (20) active physicists in non-tenure track positions (either
postdoctoral fellows, students or research assistant professors) in
the United States~\footnote{
The participant list can be found at the back of Ref.~\cite{EFTproc}.
}.  Since the meeting, three (3) of the latter
category have assumed tenure track positions, while the
remaining seventeen (17) are waiting for permanent positions
to become available to them.

\section{Other Issues}

With the large number of talented young physicists attracted to this
area of research it is important
for nuclear physics that tenure track positions (or their
equivalent) be created in the near future.  The vitality of the
field (and of nuclear physics in general) depends upon the creativity
and energy of young physicists. With  the recent exceptions of the
Jefferson Laboratory and RIKEN-BNL positions, the nuclear theory
community has had only limited success
in recruiting such talent into its ranks
and therefore a serious effort must begin now.  The alternative is
intellectual sterility.

Given that EFT has long been the {\it lingua franca} of the many
branches of particle physics, the flow of information and
understanding has largely been into nuclear physics. However,
recently, the expertise that we have developed in systems with large
scattering lengths (i.e. nuclear physics) has been applied to the
physics of BEC with great success. It is
clear that with BEC becoming only recently accessible to experimental
investigation, the overlap between the EFT program in nuclear physics
and condensed matter physics will continue to grow.  During the recent
EFT program at the {\it Institute for Nuclear Theory} in Seattle,
Washington, it became clear that the EFT tools being developed for
high precision atomic calculations will be influenced and will
influence those being developed for nuclear systems. (Incidentally,
this meeting was oversubscribed by a factor of two.)

It is also the case that realistic application of lattice-QCD results to
experimental physics requires the use of EFT methods in order to
extrapolate from the region of calculability into the regime chosen by
nature.  In addition, the confrontation of lattice-QCD (or other
theories) calculations with empirical findings at low energy
is most efficiently carried out by comparing calculated and measured
low energy constants of an EFT
rather than by dealing with matters on a process
by process basis.

Finally, it is of the utmost importance that the nuclear physics
community maintain facilities that will allow for high precision,
low-energy measurements of observables in few-nucleon systems.

\section{Acknowledgements}

This work is supported in part by the U.S. Dept. of Energy under Grants No.
DE-FG03-97ER4014 (MJS) and the National Science Foundation (BRH).

\end{document}